\documentclass[12pt]{article}
\usepackage{epsfig}

\usepackage{amssymb}
\usepackage{amsmath}
\usepackage{amsfonts}

\def\be{\begin{equation}}
\def\ee{\end{equation}}
\def\ba{\begin{array}{c}}
\def\ea{\end{array}}

\newcommand{\bea}{\begin{eqnarray}}
\newcommand{\eea}{\end{eqnarray}}

\newcommand{\kt}{\rangle}
\newcommand{\br}{\langle}
\begin{document}

\titlepage

\vspace{.35cm}

 \begin{center}{\Large \bf

A generalized family of discrete ${\cal PT}-$symmetric square wells

  }\end{center}

\vspace{10mm}

 \begin{center}

 {\bf Miloslav Znojil}

 \vspace{3mm}
Nuclear Physics Institute ASCR,

250 68 \v{R}e\v{z}, Czech Republic

{e-mail: znojil@ujf.cas.cz}

\vspace{3mm}

and

\vspace{3mm}

 {\bf Junde Wu}

 \vspace{3mm}
Department of Mathematics, College of Science, Zhejiang University,

Hangzhou 310027, Zhejiang, P. R. China

{e-mail: wjd@zju.edu.cn}


\end{center}

\vspace{5mm}

\newpage

\section*{Abstract}

$N-$site-lattice Hamiltonians $H^{(N)}$ are introduced and perceived
as a set of systematic discrete approximants of a certain ${\cal
PT}-$symmetric square-well-potential model with the real spectrum
and with a non-Hermiticity which is localized near the boundaries of
the interval. Its strength is controlled by one, two or three
parameters. The problem of the explicit construction of a nontrivial
metric which makes the theory unitary is then addressed. It is
proposed and demonstrated that due to the not too complicated (viz.,
tridiagonal matrix) form of our input Hamiltonians, the computation
of the metric is straightforward and that its matrix elements prove
obtainable, non-numerically, in elementary polynomial forms.

\section*{Keywords}

quantum mechanics; discrete lattices; non-Hermitian Hamiltonians;
Hilbert-space metrics; solvable models;

\newpage

 \section{Introduction \label{zacatek} }

{\em A priori} it is clear that the traditional and most common
physical Hilbert spaces of the admissible quantum states need not
necessarily prove optimal for computations. Once these ``obvious''
spaces ${\cal H}^{(P)}$ become distinguished by the superscript
$^{(P)}$ which may be read as an abbreviation for ``primary space'',
one may find an explicit verification of this expectation in nuclear
physics cca twenty years ago \cite{Geyer}. The amended
Schr\"{o}dinger-representation Hilbert space ${\cal H}^{(S)}$ (where
the superscript stands for ``secondary'') has been constructed there
via a  fermion-boson-space correspondence $P \leftrightarrow S$.

A perceivable simplification of the practical numerical evaluation
and/or at least of the variational prediction of the bound-state
energy levels $E_n$ has been achieved for a number of heavy nuclei.
In the notation as introduced in Ref.~\cite{SIGMA} one can identify
the underlying key mathematical idea as lying in a Dyson-inspired
ansatz connecting the $P-$superscripted and $S-$superscripted
ket-vectors,
 \be
  |\psi^{(P)}\kt\ = \ \Omega\, |\psi^{(S)}\kt \in {\cal H}^{(P)}\,,
  \ \ \ \ \ \ \
  |\psi^{(S)}\kt \in {\cal H}^{(S)}\,.
  \ee
The manipulations with the original ket vectors $|\psi^{(P)}\kt$
became, by such a construction, facilitated.

In particular, what appeared simplified was the evaluation of the
inner products $\br \phi^{(P)} |\psi^{(P)}\kt$ and of the $P-$space
matrix elements, say, of the Hamiltonian operator $\mathfrak{h}$
acting in ${\cal H}^{(P)}$. After the unitary-equivalence transition
to ${\cal H}^{(S)}$ the same quantities were represented by the new
inner products $\br \phi^{(S)} |\psi^{(S)}\kt$ and by the matrix
elements $\br \phi^{(S)}|H |\psi^{(S)}\kt$, respectively.

It is well known \cite{Dieudonne,Carl,ali} that during the
transition $P \leftrightarrow S$ between Hilbert spaces one must
also guarantee the isospectrality between the respective
Hamiltonians $\mathfrak{h}$ and $H$. In other words,, we must define
the new Hamiltonian $H$ acting in ${\cal H}^{(S)}$ by formula $H =
\Omega^{-1}\mathfrak{h}\Omega$. Then, it appears natural when the
whole change of the representation $P \to S$ is followed by another,
second-step simplification. Such a step is usually motivated by the
survival of certain cumbersome character of the work in the
secondary Hilbert space ${\cal H}^{(S)}$. In the notation of
Ref.~\cite{SIGMA}, for example, it makes sense to replace the latter
space by its ``friendlier'', auxiliary, manifestly unphysical
alternative ${\cal H}^{(F)}$.

Due to a certain freedom in the construction, the latter, third
Hilbert space may be allowed to coincide with ${\cal H}^{(S)}$ as a
topological vector space (i.e., as the space of kets,
$|\psi^{(F)}\kt| := \psi^{(S)}\kt$). What leads to the ultimate
simplicity is then the replacement of the fairly complicated,
$S-$superscripted operation ${\cal T}^{(S)}$ of the Hermitian
conjugation in ${\cal H}^{(S)}$ by the standard and trivial (i.e.,
transposition plus complex conjugation) $F-$superscripted operation
${\cal T}^{(F)}$ of the Hermitian conjugation in the final friendly
space ${\cal H}^{(F)}$.

The net purpose of the second simplification step $S \to F$ is that
the quantum system in question finds its optimal Schr\"{o}dinger
representation in ${\cal H}^{(F)}$. In this auxiliary and maximally
friendly Hilbert space one merely defines
 \be
 \br \phi^{(S)} | \psi^{(S)}\kt \ \equiv \
 \br \phi^{(F)} |\Theta | \psi^{(F)}\kt\,,
 \ \ \ \ \ \ \Theta=\Omega^\dagger\Omega\,
 \label{innerec}
 \ee
This convention keeps trace of the $S-$superscripted definition of
the physics-representing inner products in ${\cal H}^{(S)}$ and it
offers a guarantee of validity of the initial requirement of the
unitary equivalence between ${\cal H}^{(P)}$ and ${\cal H}^{(S)}$.
In a compact review~\cite{SIGMA} of the formalism we emphasized that
a given quantum bound-state system is in fact characterized by a
{\em triplet} of Hilbert spaces according to the following diagram:
 \be
  \ba
    \ \ \ \ \ \ \ \ \begin{array}{|c|}
 \hline%
  {\rm  {\bf primary,\ }difficult\  space\ P} \\
    {\rm and\ Hamiltonian}\\
  \mathfrak{h}= \mathfrak{h}^\dagger \\
 \hline
 \ea \ \ \ \ \ \ \ \ \ \ \  \  \  \ \ \ \ \ \
  \ \ \ \ \ \ \ \ \ \ \ \ \ \ \ \ \ \ \  \  \  \ \ \ \ \ \
  \ \ \ \ \ \ \ \
 \\
  \ \ \ \ \ \ \ \
  \ \ \ \ \ \ \ \
 \stackrel{{\bf Dyson \ map} \ \Omega}{}
 \ \
  \nearrow\ \  \ \ \ \ \ \ \ \
 \ \ \ \ \  \searrow \nwarrow\
 \stackrel{\bf  unitary\ equivalence}{} \ \ \ \ \ \ \ \
  \ \ \  \  \  \ \ \ \ \ \
  \ \ \ \ \ \ \ \ \ \ \ \ \ \ \ \ \ \ \  \  \  \ \ \ \ \ \
  \ \ \ \ \ \ \ \ \\
 \begin{array}{|c|}
 \hline%
  {\rm  {\bf friendly \ }but \ false\ space\ F} \\
    {\rm and\ non\!-\!Hermitian}\\
  H:=\Omega^{-1}\mathfrak{h}\Omega \neq H^\dagger \\
  \hline
 \ea
 \stackrel{ {\bf  hermitization}  }{ \longrightarrow }
 \begin{array}{|c|}
 \hline%
  {\rm  \bf secondary,\ } {\rm ultimate\ space\ S} \\
  {\rm is\ correct\ and\ physical,}\\
  H=H^\ddagger:=\Theta^{-1}H^\dagger\Theta \\
 \hline
 \ea \ \ \ \ \ \ \ \ \ \ \  \  \  \ \ \ \ \ \
  \ \ \ \ \ \ \ \ \ \ \ \ \ \ \ \ \ \ \  \  \  \ \ \ \ \ \
  \ \ \\
   \ea
 \label{THS}
 \ee
During the above-mentioned application of such a pattern to the
variational analysis of heavy nuclei it has been emphasized that,
firstly, the model itself is introduced in the P-superscripted
Hilbert space but it appeared there prohibitively complicated
\cite{Geyer}. Secondly, the successful choices of the suitable
simplification mappings $\Omega$ have been found dictated or
inspired by the underlying dynamics (i.e., in nuclei, by the
tendency of fermions to form, effectively, certain boson-resembling
clusters). Thirdly, in a way reaching far beyond the particular
nuclear physics context, the product $\Omega^\dagger\Omega = \Theta
\neq I$ has been noticed to play the role of the metric in the
ultimate, S-superscripted Hilbert-space.

Cca ten years ago, the metric-operator interpretation of nontrivial
$\Theta \neq I$ became believed to apply to a very broad family of
models including, typically, the imaginary-cubic oscillator
 \be
 H=-\frac{d^2}{dx^2}+{\rm i}x^3
 \label{icu}
 \ee
as well as many other Hamiltonians $H$ introduced as acting in
${\cal H}^{(F)}:=L^2(\mathbb{R})$ and/or in ${\cal H}^{(S)} \neq
L^2(\mathbb{R})$ and relevant, typically, in the relativistic
quantum field theory (cf., e.g., \cite{Carl} or \cite{ali} for
extensive details).

The basic ideas behind the pattern of Eq.~(\ref{THS}) were broadly
accepted and the whole mathematical formalism (which we call,
conveniently, the three-Hilbert-space (THS) representation of
quantum states) started to be treated as an old and well understood
one. In the year 2012, this opinion has rather drastically been
challenged by the results of Refs.~\cite{Siegl} where it has been
proved, {\em rigorously}, that for the most popular ``benchmark''
THS model (\ref{icu}) the class of the eligible Hilbert-space metric
operators $\Theta$ is in fact {\em empty}. In other words we were
all suddenly exposed to the necessity of reanalyzing the mathematics
behind the differential-operator models as sampled by
Eq.~(\ref{icu}).

This observation belongs to one of the key motivations of our
present study. The emergence of incompatibility of the overall
methodical THS pattern (\ref{THS}) with the concrete
unbounded-operator example~(\ref{icu}) implies that the attention of
mathematical physicists must immediately be redirected and returned
to the alternative, mathematically correct benchmark models like,
e.g., the bounded-operator Hamiltonians of Ref.~\cite{Geyer} and/or
even to the most schematic, exactly solvable finite-dimensional
models as sampled, say, by the non-numerical discrete square well of
our preceding Paper 1 \cite{I}.

The latter family of models was characterized by the sequence of the
most elementary finite-dimensional Hamiltonians
 $$
H^{(3)}({\lambda})=\left[ \begin {array}{ccc}
2&-1-{{\lambda}}&0\\{}-1+{{\lambda}}&2&-1+{{\lambda}}
\\{}0&-1-{{\lambda}}&2\end
{array} \right]\,,
 $$
 $$
 H^{(4)}({\lambda})=\left[ \begin {array}{cccc} 2&-1-{\it
{\lambda}}&0&0\\{}-1+{ \it {\lambda}}&2&-1&0\\{}0&-1&2&-1+{\it
{\lambda}}
\\{}0&0&-1-{\it {\lambda}}&2\end {array} \right]\,
 $$
 $$
  H^{(5)}({\lambda})=\left[ \begin {array}{ccccc} 2&-1-{\it
{\lambda}}&0&0&0\\{}-1+{\it {\lambda}}&2&-1&0&0\\{}0&-1&2&-1&0\\{}0
&0&-1&2&-1+{\it {\lambda}}\\{}0&0&0&-1-{\it {\lambda}}&2\end {array}
 \right]
 $$
i.e., by the matrix
 \be
  H^{(N)}({\lambda})=  \left[ \begin {array}{cccccc}
 2&-1-{\it {\lambda}}&0&\ldots&0&0
\\
{}-1+{\it {\lambda}}&2&-1&0&\ldots&0
\\
{}0&-1&\ \ \ 2\ \ \ &\ddots&\ddots&\vdots
\\
{}\vdots&0&\ddots&\ \ \ \ddots\ \ \ &-1&0
\\
{}0&\vdots&\ddots&-1&2&- 1+{\it {\lambda}}
\\
{}0&0&\ldots&0&-1-{\it {\lambda}}&2
\end {array}
 \right]\,
 \label{toym}
 \ee
considered at an arbitrary preselected Hilbert-space dimension $N$.
As required, this matrix appears non-Hermitian in the
$N-$dimensional and manifestly unphysical, auxiliary (and, in our
case, real) Hilbert space ${\cal H}^{(F)}_{(N)} \ \equiv \
\mathbb{R}^N$ where the inner product remains trivial,
 $$
 \br \phi^{(F)} |\psi^{(F)}\kt = \sum_{n=1}^N\
 \phi^{(F)}_n\,\psi^{(F)}_n\,.
 $$
In Paper 1 we emphasized that one may try to deduce the physical
context, contents and meaning of models (\ref{toym}) in their $N\to
\infty$ limiting coincidence with certain usual single-parametric
differential Schr\"{o}dinger operators on the line \cite{david}.

In the additional, methodical role of non-contradictory and exactly
solvable, non-numerical benchmark models, the most serious weakness
of Hamiltonians (\ref{toym}) may be seen in their trivial
kinetic-operator nature inside the whole interior of the interval of
the spatial coordinate $x$ (see also Paper 1 for a more explicit
explanation and further references). This means that their
nontrivial dynamical content (i.e., their point-like-interaction
component) is  merely one-parametric and restricted to the points of
the spatial boundary.

In our present paper we intend to extend this perspective in a
systematic manner by showing, first of all, that the latter weakness
of the models of Paper 1 is curable. We shall introduce and employ a
few less elementary toy-model interactions on the same $N-$site
quantum lattice. In Section \ref{jedenpar} we select just a less
trivial version of the one-parametric interaction while in
subsequent Sections \ref{dvapar} and \ref{tripar}, two and three
parameters controlling the interaction are introduced, respectively.
Our overall message is finally summarized in Sections \ref{outl}
(discussion and outlook) and \ref{summary} (summary).

\section{A slightly more sophisticated one-parametric model\label{jedenpar}}

\subsection{Hamiltonians $H^{(N)}$ and metrics $\Theta^{(N)} $}

Let us consider a non-Hermitian and real $N$ by $N$ Hamiltonian
matrix ${H}$ in which the interaction connects the triplets of the
next-to-the-boundary sites,
 \be
 {H}^{(N)}({\lambda})=\left[
 \begin {array}{cccccccc}
     2&-1-{\it {\lambda}}&0&0&\ldots&&\ldots&0
 \\{}
 -1+{\it {\lambda}}&2&-1+{\it {\lambda}}&0&\ldots&&&\vdots
 \\{}
 0&-1-{\it {\lambda}}&2&-1&\ddots&&&
 \\{}
 0&0&-1&2&\ddots&\ddots&\vdots&\vdots
 \\{}
  \vdots&\vdots&\ddots&\ddots&\ddots&-1&0&0
 \\{}
  &&&\ddots&-1&2&-1-{\it {\lambda}}&0
 \\{}
 \vdots&&&\ldots&0&-1+{\it {\lambda}}&2&-1+{\it {\lambda}}
 \\{}
 0&\ldots&&\ldots&0&0&-1-{\it {\lambda}}&2
 \end {array}
 \right]\,.
 \label{ham11}
 \ee
Recalling the experience gained in Paper 1 we may expect that the
bound-state-energy eigenvalues obtained from this Hamiltonian will
be all real at the sufficiently small values of the couplings
$\lambda \in (-a,a)$ with, presumably, $a=1$.

A rigorous proof of the above conjecture would be feasible albeit
lengthy. Although we are not going to present it here due to the
lack of space, Figure \ref{bigthee} samples the whole spectrum at
$N=11$ and offers a persuasive numerical support of such an
expectation. Moreover, a comparison of this picture with its
predecessors of Paper 1 indicates that the use of a less trivial
Hamiltonian seems truly rewarding. In the past, the
phenomenologically rich and promising nontrivial structure of the
parameter-dependence of the spectrum near $\lambda \approx a$
motivated quite strongly the continuation of our study of similar,
more complicated toy models.


\begin{figure}[h]                     
\begin{center}                         
\epsfig{file=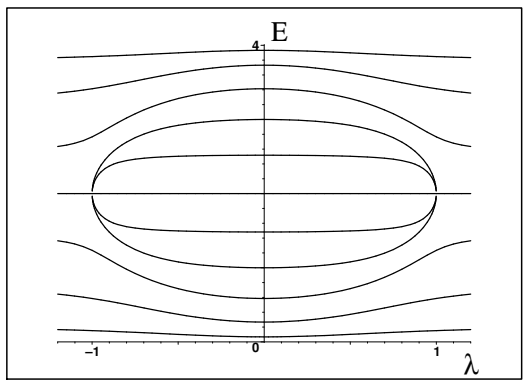,angle=270,width=0.5\textwidth}
\end{center}                         
\vspace{-2mm}\caption{The $\lambda-$dependence of the eigenvalues of
Hamiltonian (\ref{ham11}). Obviously, this spectrum stays real in
the interval of $\lambda\in (-1,1)$.}
 \label{bigthee}
\end{figure}

Under the hypothesis of the reality of the spectrum, a completion of
the construction of the corresponding consistent THS quantum model
requires, naturally, the explicit construction of a metric $\Theta$
entering the physical inner product (\ref{innerec}). In its full
completeness, such a task has been pursued in Paper 1. In what
follows we intend to complement this research towards some more
complicated Hamiltonians sampled by Eq.~(\ref{ham11}) above. At the
same time we shall skip all details of an exhaustive analysis and
reduce the exhaustive constructive classification of the
$N-$parametric sets of metrics
 \be
   \Theta^{(N)}=
   \sum_{k=1}^N\, {\mu_k}\,{\cal P}^{(k)}\,
 \label{e777}
   \ee
to the mere evaluation of a characteristic sample of its individual
Hermitian-matrix components ${\cal P}_k^{(N)}$. These components may
be interpreted as metric-resembling (i.e., not necessarily positive
definite) matrices. Their main pedagogical merit is that they remain
sufficiently transparent matrices with, hopefully, sparse structure
of the universal form which has been found and described in Paper 1.

With this purpose in mind we shall require that the individual
components of the sum Eq.~(\ref{e777}) satisfy the Dieudonn\'{e}
equation {\em alias} quasi-Hermiticity condition
 \be
 \sum_{m=1}^N\,
 \left [
      \left (H^\dagger\right )_{jm}\,{\cal P}_{mn}
      -{\cal P}_{jm}\,H_{mn}\right ] =0
 \,,\ \ \ \ \ j,n=1,2,\ldots,N
   \,.
 \label{htot}
 \ee
In the light of the analysis of Paper 1 we shall, furthermore, save
time and skip the exhaustive discussion of the (more or less
trivial) general $N-$dependence of the model. In order to gain an
overall insight into the structure of the THS representability of
our model, we found it sufficient to restrict attention to a fixed
value of dimension $N$ which is neither too small (we have to avoid
the structural degeneracies at small $N$) nor too large (we intend
to display some matrices in print).

\subsection{Matrix ${\cal P}^{(6)}$ at $N=11$}

Following the recipe described in Paper 1 we shall start from the
ansatz
 \be
  {\cal P}^{(6)}=\left[ \begin {array}{ccccccccccc} 0&0&0&0&0&r&0&0&0&0&0
\\{}0&0&0&0&s&0&s&0&0&0&0\\{}0&0&0&v&0
&t&0&v&0&0&0\\{}0&0&v&0&w&0&w&0&v&0&0
\\{}0&s&0&w&0&1&0&w&0&s&0\\{}r&0&t&0&1
&0&1&0&t&0&r\\{}0&s&0&w&0&1&0&w&0&s&0
\\{}0&0&v&0&w&0&w&0&v&0&0\\{}0&0&0&v&0
&t&0&v&0&0&0\\{}0&0&0&0&s&0&s&0&0&0&0
\\{}0&0&0&0&0&r&0&0&0&0&0\end {array} \right]
\label{anza}
 \ee
and, in the light of Eq.~(\ref{htot}), we shall compare the matrix
product ${\cal P}^{(6)}\,H$
%
with the matrix product $H^\dagger{\cal P}^{(6)}$.
%
%
Element by element, their (row-wise running) comparison yields the
nontrivial constraints $s= s {\lambda} + r$ in the fifth and seventh
step, $v= - v {\lambda} + s$ in the fifteenth step, etc. After the
tedious though entirely straightforward manipulations we obtain the
final solution/formulae
  $$
  r={\frac {1-{{\it {\lambda}}}^{2}}{1+3\,{{\it {\lambda}}}^{2}}}\,,\ \ \
  s={\frac {1+{\it {\lambda}}}{1+3\,{{\it {\lambda}}}^{2}}}\,,\ \ \
  v={\frac {1}{1+3\,{{\it {\lambda}}}^{2}}}
  $$
  \be
  t={\frac {1+{{\it {\lambda}}}^{2}}{1+3\,{{\it {\lambda}}}^{2}}}\,,\ \ \
  w={\frac {1+2\,{{\it {\lambda}}}^{2}}{1+3\,{{\it {\lambda}}}^{2}}}
  \label{reza}
  \ee
which indicate that the transition to the more-site interactions in
the Hamiltonian may still be expected to lead to the polynomial or
rational-function dependence of the matrix elements of the metric on
the value of the coupling constant. The second, methodically equally
encouraging consequence of the construction of the sample
pseudometric ${\cal P}^{(6)}$ is that after a not too drastic loss
of the simplicity of the input matrix Hamiltonians the construction
of the class of admissible metric remains feasible by non-numerical
means. Thirdly, via a deeper analysis of Dieudonn\'{e}'s Eq.
(\ref{htot}) it is easy to deduce that the extension of the $N=11$
results to any dimension $N> 11$ parallels the pattern found in
Paper 1 and degenerates to a virtually trivial extrapolation of the
interior parts of individual items  ${\cal P}^{(k)}$ in the matrix
sequences determining the general metric (\ref{e777}).

%

\section{Two-parametric Hamiltonians \label{dvapar}}

\subsection{Energies}

%
%
%
%
%
%
%
%

Once we recall preceding section and disentangle the values of the
respective couplings between the two next-to-boundary and two
next-to-next-to-boundary sites we obtain the following
two-parametric $N=11$ Hamiltonian matrix
 \be
 \left[ \begin {array}{ccccccccccc} 2&-1-{\it {\lambda}}&0&0&0&0&0&0&0&0&0
\\{}-1+{\it {\lambda}}&2&-1+{\it {\mu}}&0&0&0&0&0&0&0&0
\\{}0&-1-{\it {\mu}}&2&-1&0&0&0&0&0&0&0
\\{}0&0&-1&2&-1&0&0&0&0&0&0\\{}0&0&0&-
1&2&-1&0&0&0&0&0\\{}0&0&0&0&-1&2&-1&0&0&0&0
\\{}0&0&0&0&0&-1&2&-1&0&0&0\\{}0&0&0&0
&0&0&-1&2&-1&0&0\\{}0&0&0&0&0&0&0&-1&2&-1-{\it {\mu}}&0
\\{}0&0&0&0&0&0&0&0&-1+{\it {\mu}}&2&-1+{\it {\lambda}}
\\{}0&0&0&0&0&0&0&0&0&-1-{\it {\lambda}}&2\end {array}
 \right]
 \label{ham11be}
 \ee
Its full display still almost fits in the printed page but what is
certainly more important is that the presence of the new variable
coupling $\mu$ extends the capability of the model of being more
useful in some phenomenologically oriented considerations. This
seems well illustrated by Fig.~\ref{thee2} where we restricted
attention to a line in the plane of parameters defined by the
constraint $\mu \to \mu(\lambda)= \lambda+{\rm a\ constant}$.


\begin{figure}[h]                     
\begin{center}                         
\epsfig{file=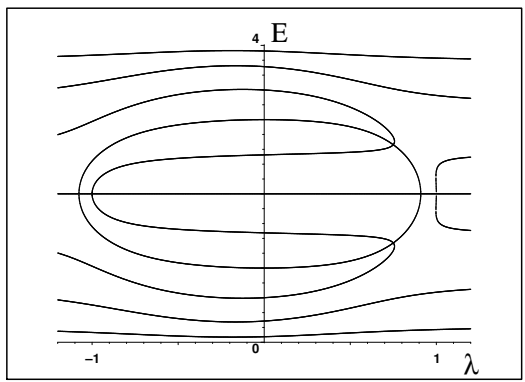,angle=270,width=0.5\textwidth}
\end{center}                         
\vspace{-2mm}\caption{The $\lambda-$dependence of the eigenvalues of
Hamiltonian (\ref{ham11be}) in which we selected the constantly
shifted value of $\mu =\mu(\lambda):= \lambda+0.25$.}
 \label{thee2}
\end{figure}

One can easily check, in Fig.~\ref{thee2}, that the original picture
lost its left-right symmetry and that the real bound-state-energy
values only occur in a smaller, asymmetric interval of $\lambda \in
(-1,b)$ where $b \approx 0.75$ for our particular illustrative
choice of the constant shift $\Delta=\mu-\lambda$. The further
inspection of the picture reveals many further and qualitatively
interesting features of the ``phase transition'' during which the
pairs of individual energy levels cross or merge and, subsequently,
complexify. Temporarily, some of the complexified pairs may even
return to the reality later -- notice, in the picture, that there
are as many as nine real level at $\lambda$s which lie slightly
below the critical $\lambda=1$.

\subsection{Pseudometrics}

In a way paralleling the preceding section we shall now restrict
attention to the intervals of $\lambda \in (a(\mu),b(\mu))$ and $\mu
\in (c(\lambda),d(\lambda))$, i.e., to the two-dimensional physical
domain ${\cal D}$ of ``acceptable'' parameters in the Hamiltonian.
Inside this domain the whole spectrum remains, by definition,
completely real and non-degenerate, i.e., potentially, physical,
observable and compatible with the unitarity of the time evolution.

In this setting the obligatory construction of the suitable matrices
of the metric may proceed along the same lines as above. In full
parallel, we shall therefore return to the independent variability
of the two couplings in the Hamiltonian and reopen the problem of
the construction of the metric via Eq.~(\ref{e777}). In the language
of Ref. \cite{Geyer}, the $N-$parametric ambiguity contained in the
latter formula makes the related picture of physics flexible and
adaptable to our potential choice of further relevant operators
(i.e., in our case, of some $N$ by $N$ matrices) of observables.

Under our present restricted project, we shall again pay attention
merely to the explicit construction of the ``most interesting'' $N$
by $N$ pseudometric ${\cal P}^{(J)}$ at $J=6$ and $N=2J+1=13$. The
method of construction will remain the same. During its application
we displayed, first of all, the non-vanishing matrix elements of the
sparse difference matrix $H^\dagger{\cal P}^{(6)}-{\cal P}^{(6)}\,H$
and made them equal to zero via the solution of the corresonding
algebraic equations.
%
%
%
At the end of this procedure which completely paralleled our
preceding use of ansatz (\ref{anza}) as well the format of result
(\ref{reza}) we obtained the matrix elements of our sample
pseudometric ${\cal P}^{(6)}$ in the following, equally compact and
comparably transparent form
 $$
 r={\frac { \left( 1+{\it {\mu}} \right)
  \left( 1-{\it {\lambda}} \right) }{1+{{
 \it {\lambda}}}^{2}+2\,{{\it {\mu}}}^{2}}}\,, \ \ \ \ \
 s={\frac
 {1+{\it {\mu}}}{1+{{\it {\lambda}}}^{2}+2\,{{\it {\mu}}}^{2}}}\,, \ \ \ \ \
 v={\frac {1}
 {1+{{\it {\lambda}}}^{2}+2\,{{\it {\mu}}}^{2}}}
 $$
 $$
 t={\frac {1+{{\it {\lambda}}}^{2}}{1+{{\it {\lambda}}}^{2}
 +2\,{{\it {\mu}}}^{2}}}\,, \ \ \ \ \
 w={\frac {1+{{\it {\lambda}}}^{2}+{{\it {\mu}}}^{2}}
 {1+{{\it {\lambda}}}^{2}+2\,{{\it {\mu}}}^{2}}}\,.
 $$
One should add here that due to the multiple symmetries of our
Hamiltonian matrix as well as of the metric, the inversion of the
metric (or pseudometric) may be obtained by the simple change of the
sign of the pair of our coupling-constant quantities ${\lambda}$ and
${\mu}$. The inspection of the latter formulae also reveals that the
numerators remain the same so that they might be all omitted or
ignored as an inessential overall multiplication factor.

\section{Three-parametric Hamiltonians\label{tripar}}

For a proper, non-degenerate tractability of the next family of some
three-parametric Hamiltonians we need to deal with the dimensions
$N\geq 13$ at least. The full matrices will not fit in the printed
page anymore. Fortunately, their numerous symmetries will still
allow us to display the relevant information about their matrix
elements. In particular, it proves sufficient to display just the
upper part of the Hamiltonian matrix in full detail,
%
 $$
 H^{(N)}=\left[ \begin {array}{ccccccc}
  2&-1-{\it {\lambda}}&0&\ldots&&\ldots&0
 \\{}-1+{\it {\lambda}}&2&-1+{\it {\mu}}&0&\ldots&&\vdots
 \\{}0&-1-{\it {\mu}}&2&-1-{\nu}&0&\ldots&
 \\{}\vdots&0&-1+{\nu}&2&-1&\ddots&\vdots
 \\{}&&\ddots&-1&2&\ddots&0
  \\{}\vdots&&&\ddots&\ddots&\ddots&-1+{\it {\lambda}}
 \\{}0&\ldots&\ldots&0&0&-1-{\it {\lambda}}&2\end {array}
 \right]\,.
 $$
Similarly, the symmetries of the most interesting $N=13$
pseudometric component ${\cal P}^{(7)}$ of the $N=13$ metric
(\ref{anza}) enables us to search for its matrix elements via the
thirteen-dimensional matrix ansatz
\newpage
 \be
  {\cal P}^{(7)}=\left[ \begin {array}{ccccccccccccc}
     0&\ldots&&&\ldots&0&r&0&\ldots&&&\ldots&0
 \\{}\vdots&&&\ldots&0&s&0&s&0&\ldots&&&\vdots
 \\{}&&\ldots&0&p&0&t&0&p&0&\ldots&&
 \\{}&& {\large \bf _. } \cdot {\large \bf ^{^.}}
 &v&0&q&0&q&0&v&\ddots&&
 \\{}&
 & {\large \bf _. } \cdot {\large \bf ^{^.}}
 &0&w&0&m&0&w&0&\ddots&&
 \\{}&
 &{\large \bf _. } \cdot {\large \bf ^{^.}}
 & {\large \bf _. } \cdot {\large \bf ^{^.}}
 &0&u&0&u&0&\ddots&\ddots&&
 \\{}
 &
 &
 & {\large \bf _. } \cdot {\large \bf ^{^.}}
 & {\large \bf _. } \cdot {\large \bf ^{^.}}
 &0&1&0&\ddots&\ddots&&&
 \\{}\vdots&
 &
 &
 &
 &
 &\vdots&&
 &&&&\vdots
 \\
  0&\ldots&&\ldots&0&0&r&0&0&\ldots&&\ldots&0
 \end {array} \right]
 \label{anzabr}
 \ee
It is worth adding that wherever we decide to choose $N > 13$, the
triple dots may be read here as indicating, for all of the sharply
larger dimensions, simply the repetition of the same (i.e., of the
last) element until the symmetry of the matrix allows.

Strictly the same procedure as above leads again to the final and
still amazingly compact solution
  $$
  r={\frac { \left( 1-{\it {\nu}} \right)  \left( 1+{\it {\mu}} \right)
   \left(
1-{\it {\lambda}} \right) }{1+{{\it {\lambda}}}^{2}+2\,{{\it
{\mu}}}^{2}+3\,{{\it {\nu}}}^ {2}+{{\it {\nu}}}^{2}{{\it
{\lambda}}}^{2}}}
 \,, \ \ \ \ \ \
 s={\frac { \left( 1-{\it {\nu}} \right)  \left( 1+{\it {\mu}} \right) }{1+{{
\it {\lambda}}}^{2}+2\,{{\it {\mu}}}^{2}+3\,{{\it {\nu}}}^{2}+{{\it
{\nu}}}^{2}{{\it {\lambda}}}^{2}}}
 $$
 $$
 p={\frac {1-{\it {\nu}}}{1+{{\it {\lambda}}}^{2}+2\,{{\it {\mu}}}^{2}
 +3\,{{\it {\nu}}}^{
2}+{{\it {\nu}}}^{2}{{\it {\lambda}}}^{2}}}
 \,, \ \ \ \ \ \
 v={\frac {1}{1+{{\it {\lambda}}}^{2}+2\,{{\it {\mu}}}^{2}+3\,{{\it {\nu}}}^{
2}+{{\it {\nu}}}^{2}{{\it {\lambda}}}^{2}}}
 $$
 $$
 t= {\frac { \left( 1-{\it {\nu}} \right)   \left( 1+{\it {\lambda}}^2
 \right) }{1+{{\it {\lambda}}}^{2}+2\,{{\it {\mu}}}^{2}
 +3\,{{\it {\nu}}}^{2}+{{\it
{\nu}}}^{2}{{\it {\lambda}}}^{2}}}
 \,, \ \ \ \ \ \
 q={\frac {1+{{\it {\mu}}}^{2}+{{\it {\lambda}}}^{2}}
 {1+{{\it {\lambda}}}^{2}+2\,{{\it
{\mu}}}^{2}+3\,{{ \it {\nu}}}^{2}+{{\it {\nu}}}^{2}{{\it
{\lambda}}}^{2}}}
 $$
 $$
 w={\frac {1+{{\it {\mu}}}^{2}+{{\it {\nu}}}^{2}
 +{{\it {\lambda}}}^{2}}{1+{{\it {\lambda}}}^{
2}+2\,{{\it {\mu}}}^{2}+3\,{{\it {\nu}}}^{2}+{{\it {\nu}}}^{2}{{\it
{\lambda}}}^{2}}}
 \,, \ \ \ \ \ \
 m=1-2\,{\frac {{{\it {\nu}}}^{2}}{1+{{\it {\lambda}}}^{2}
 +2\,{{\it {\mu}}}^{2}+3\,{{
\it {\nu}}}^{2}+{{\it {\nu}}}^{2}{{\it {\lambda}}}^{2}}}
 $$
 $$
 u=1-{\frac {{{\it {\nu}}}^{2}}{1+{{\it {\lambda}}}^{2}
 +2\,{{\it {\mu}}}^{2}+3\,{{\it
{\nu}}}^{2}+{{\it {\nu}}}^{2}{{\it {\lambda}}}^{2}}}
 $$
From this set of formulae we may extract the similar messages as
above.

\section{Discussion \label{outl}}

In the sense of commentaries scattered over the preceding sections
we now intend to complement the preceding Summary section by an
outline of a few possible future mathematical and methodical as well
as purely phenomenologically motivated extensions of the model.

In the corresponding list of the possible directions of a
generalization of the present model, the one which looks most worth
pursuing lies in the systematic search for the further exactly
solvable finite-dimensional models which would admit not only the
closed-form representation of the real spectrum of the energies but
also the explicit construction of the metric operator. Even if one
would be able to construct just some (i.e., not all) metrics (which
is, after all, most common in the literature), the scarcity of the
exactly solvable models in this field would certainly provide a
ground for the publication of this type of the results.

By our recommendation one might particularly concentrate attention
to the preservation of the localized support of the interactions
near the corners of the tridiagonal Hamiltonian matrix. This idea
was originally inspired by the discovery of the tractability of the
differential-equation $N \to \infty$ models with point interactions
at the boundaries \cite{david}. At the finite dimensions, the same
features of the dynamics have now been found to survive even in the
models constructed at the not too large dimensions $N \ll \infty$.
We believe, therefore, that the latter choice of the specific
dynamics will gain further popularity as a ground of an optimal
solvable-model-building strategy in the nearest future.

Certainly, there exist further interesting aspects of a systematic,
model-based quantum mechanics of the elementary models which look
non-Hermitian when solely considered in the most user-friendly,
F-superscripted Hilbert space ${\cal H}^{(F)}$. One of the most
obvious apparent paradoxes may be seen in the mathematical
non-uniqueness of the assignment of the metric $\Theta$ to a given
Hamiltonian $H$. Fortunately, the answer has already been provided
twenty years ago when the authors of  Ref. \cite{Geyer} gave the
complete answer. Briefly stated: the ambiguity $\Theta=\Theta(H)$
merely reflects the open possibility of incorporation of additional
phenomenological information via an introduction of more observable
quantities.

The best known illustrative example of such an added observable is
the Bender's ``charge'' \cite{Carl}. Now, whenever one chooses this
charge or another observable as a phenomenological input, the
possibility and feasibility of the construction of the complete
family of the eligible metrics $\Theta=\Theta(H)$ in a closed,
non-numerical form will always represent a significant advantage of
the mathematical model. Plus, needless to add, the use of any
analytic though still flexible form of the metric which appears in
the mean values, i.e., in principle, which enters all of the
measurable predictions would certainly enhance the appeal of the
theory in applications.

Another apparent paradox concerns the ``kinematical'' multi-index
parameter $\alpha$ which reflects the above-mentioned ambiguity and
which numbers the alternative eligible metrics $\Theta(H)
=\Theta_\alpha (H)$. It is obvious that for some values
$\alpha_{critical}$ of these parameters the metric itself may become
singular and unacceptable. An interesting potential reward of the
further study of a particular quantum model characterized by an
operator (or, in our case, matrix) doublet ($H,\Theta_\alpha$) might
be seen in the possible quantitative specification of the
connections between the critical values of $\alpha_{critical}$ as
functions, say, of the (possibly, multi-index) dynamics-determining
couplings $\lambda$ in $H=H(\lambda)$.

In some sense, the closely related and/or complementary questions
will also emerge in connection with any toy-model $H=H(\lambda)$ in
which the complexification of the eigenenergies occurs at the so
called Kato's exceptional points $\lambda_{cricical}$ (at which the
energies merge and subsequently complexify - for illustration see,
e.g., the presence of the pair of exceptional points
$\lambda_{cricical}=\pm 1$ in Figure \ref{bigthee}). In particular,
an explicit future construction of solvable models might be able to
clarify the mutual connections between, firstly, the ``dynamical''
loss of the observability of the energies at
$\lambda=\lambda_{cricical}$ and, secondly, the ``kinematical'' loss
of the existence of the pre-selected S-superscripted Hilbert space
at $\alpha=\alpha_{cricical}$ in connection, thirdly, with the
necessary loss of the observability of some {\em other} dynamical
observable at {\em the same} $\alpha=\alpha_{cricical}$ (the readers
should consult, first of all, Ref.~\cite{Geyer} in this context).

Last but not least, another natural future continuation of research
which may be expected exceptionally promising might concentrate upon
the scenario in which the eigenvalues of $H$ remain real while the
metric re-regularizes ``insufficiently'', becoming merely indefinite
after the parameter $\alpha$ itself crosses, in an appropriate
manner, the critical value of $\alpha_{cricical}$. In such a
context, one might merely re-classify the resulting ``wrong'' or
``indefinite'' metric $\Theta_\alpha$ as the Bender's ``parity''
${\cal P}$ and search for his ``charge'' ${\cal C}$ in the ``new
metric'' $\Theta_{changed}={\cal PC}$ (cf. \cite{Carl} for the
complete recipe).

\section{Summary\label{summary}}

On the background of comparison with the older results of Paper 1,
one of the most surprising features of their present generalization
may certainly be seen in the friendly nature of the more-parametric
formulae. A completion and further extension of such constructions
along the lines indicated in preceding section seems to be a project
with good chances for a success in the future, indeed.

Our present first results in this direction may be briefly
summarized as follows. Firstly, we revealed an emergent pattern of
having, up to an overall factor, the purely polynomial matrix
elements of the ``pseudometric'' components ${\cal P}$ of the
metrics. Our sample calculations found such a hypothesis
reconfirmed.

Secondly, we may feel impressed by the emergence of the pattern of
most natural and obvious further generalizations of the Hamiltonians
in which one introduces new and new  parameters at an increasing
distance from the boundaries of the lattice. It is certainly
encouraging that such a recipe leaves the construction non-numerical
and that it seems to offer unexpectedly compact and transparent
benchmark-type results.

\section*{Acknowledgments}

Participation of MZ was supported by the GA\v{C}R grant Nr.
P203/11/1433. Participation of JW was supported by the Natural
Science Foundations of China (11171301) and by the Doctoral Programs
Foundation of Ministry of Education of China (J20130061).

\end{document}